\documentclass[%
 reprint,
 amsmath,amssymb,
 aps,
floatfix,
]{revtex4-2}
\usepackage{float}
\usepackage{soul, xcolor}
\usepackage{parskip}
\usepackage{graphicx}
\usepackage{dcolumn}
\usepackage{bm}

\begin{document}
\setstcolor{red}
\preprint{APS/123-QED}

\title{Spin wave excitation and directional propagation in presence of magnetic charges in square artificial spin ice }
\author{Nimisha Arora}
\email{narora1.physics@gmail.com}
\author{Pintu Das}
\email{pdas@physics.iitd.ac.in}
\affiliation{Department of Physics, Indian Institute of Technology Delhi, New Delhi, India}
\begin{abstract}
Artificial spin ice is a special class of engineered lattice of highly shape anisotropic single domain magnetic nanostructures which is used as one of the model systems to study the spin ice behavior observed in pyrochlore oxides. The nanomagnets interact via dipolar interaction which results in correlated magnetization dynamics exhibiting macroscopic spin configuration states. 
Here, we exploit the interplay of underlying magnetic state and external bias field orientation to study controlled spin wave propagation in square Artificial Spin Ice (sASI) by performing detailed micromagnetic simulations. We report that careful selection of vertices with local magnetic charges can effectively direct the anisotropic spin wave in presence of an external field. Further, we explore the influence of local charges due to the excited state in even-coordinated vertices as well as uncompensated charges due to odd-coordinated vertices on spin wave behavior. Our studies suggest that there is no perceptible difference on spin wave dynamical behavior due to the origin of local magnetic charge in sASI. Our results of controlled and directional spin wave propagation in sASI system may be useful for low-power consumption based all magnonic on-chip devices.

\end{abstract}

\maketitle

\section{\label{sec:level1}Introduction}
As the conventional electronic charge transport based technological devices approaching the quantum limit, there is a growing thrust on the energy efficient and faster devices based on electron's spin. 
Recent development has shown that collective precession of localized magnetic moments (excitation of spin waves) can be used to carry information to a much larger distance in comparison to the conduction electron's spin current. The quantum of spin waves (SWs) is called magnon and the field of studying SWs in magnetic nanostructures has emerged under the name of magnonics~\cite{kruglyak2010preface}.  
From a classical perspective, SWs can be treated as perturbations in phase-coherent precession of magnetization vectors, which propagate in coupled magnetic media. Due to the very short wavelengths of SWs (1\,$\mu$m to 150\,nm depending on the lattice constant of the magnetic material), these are emerging as potential information carriers in energy efficient miniaturized spin based devices in the nanoscale regime. Thus, excitation of SWs of tuned wavelength is one of the major topics of current research in this area\,\cite{neusser2009magnonics,kruglyak2010preface,lenk2011building,demokritov2012magnonics, nikitov2015magnonics, liu2018organic}. The other major considerations for the SWs based devices are the propagation distance, group velocity, phase velocity of the SWs, etc.\,\cite{neusser2009magnonics,kruglyak2010preface,lenk2011building,demokritov2012magnonics, nikitov2015magnonics, liu2018organic}. Several of these parameters depend on the competition between the exchange and dipolar interactions of the spins. Recent studies have reported several SW based functional device applications in phase shifters\,\cite{au2012nanoscale}, microwave antenna\,\cite{abeed2020experimental}, directional coupler\,\cite{wang2018reconfigurable}, SW-based multiplexer\,\cite{vogt2014realization}, interferometer\,\cite{kostylev2005spin, papp2017nanoscale, chen2021reconfigurable}, grating\,\cite{yu2013omnidirectional}, neuromorphic computing\,\cite{grollier2016spintronic, arava2018computational, arava2019engineering, torrejon2017neuromorphic}, and memory register\,\cite{caravelli2022artificial} etc. Furthermore, it has been reported that phase incoherent electromagnetic wave can give rise to multiple phase-coherent magnons, which shows the application in energy harnessing and information transfer from alternative sources\,\cite{li2019strong}.
Thus, it is clear that successful implementation of magnonics in modern devices requires the SWs to be controllably guided through nanostructures. In order to acquire this ability, fundamental and in-depth understanding of the behavior of SWs in magnetic materials of different nanoscale geometries is very important. Recently a few experimental as well as theoretical results on the investigation of this aspect of SWs have been reported in literature~\cite{kaffash2021nanomagnonics, lendinez2021observation, gartside2021reconfigurable, caravelli2020logical, iacocca2020tailoring, gliga2020dynamics}. The magnetic nanostructures, which are involved in the propagation of SWs, may in general interact via long range dipolar interactions. Therefore, such magnetostatic interaction may play a role on the propagation of SWs in dipolarly coupled magnetic nanostructures which is far from clearly understood. This clearly underlines the importance of investigation of general behavior of SWs under dipolarly coupled environment.\\ 
In this work, we have studied the behavior of excited SWs in highly shape anisotropic nanomagnets of Ni$_{80}$Fe$_{20}$ arranged in the sASI geometry~\cite{wang2006artificial}. Typically, these designer nanostructured materials offer alluring possibilities to study physics of geometric frustration~\cite{wang2006artificial}, emergent magnetic monopoles~\cite{morgan2011thermal, keswani2021controlled}, collective magnetization dynamics of interacting spins~\cite{jungfleisch2017high}, and phase transition~\cite{kapaklis2012melting}, etc. at room temperature while offering the possibility of visualizing individual macrospins. Owing to the possibility of controlling magnetic microstates by designing different structures such as kagome~\cite{qi2008direct}, sASI~\cite{wang2006artificial}, etc. where dipolar interactions play a significant role in defining the exact magnetic microstates, these nanoscale magnetic structures are interesting candidates to study the interplay of dipolar interaction and SW propagation. Although, recently some attention has been directed towards the SW behavior of ASI systems~\cite{talapatra2021coupled, chaurasiya2021comparison, saha2021spin, arora2021spin, arroo2019sculpting, iacocca2016reconfigurable, montoncello2018mutual}, the role of dipolar interactions and associated magnetic charges on the SW behavior has not been investigated in details.\\
\begin{figure}
\includegraphics[width=\linewidth]{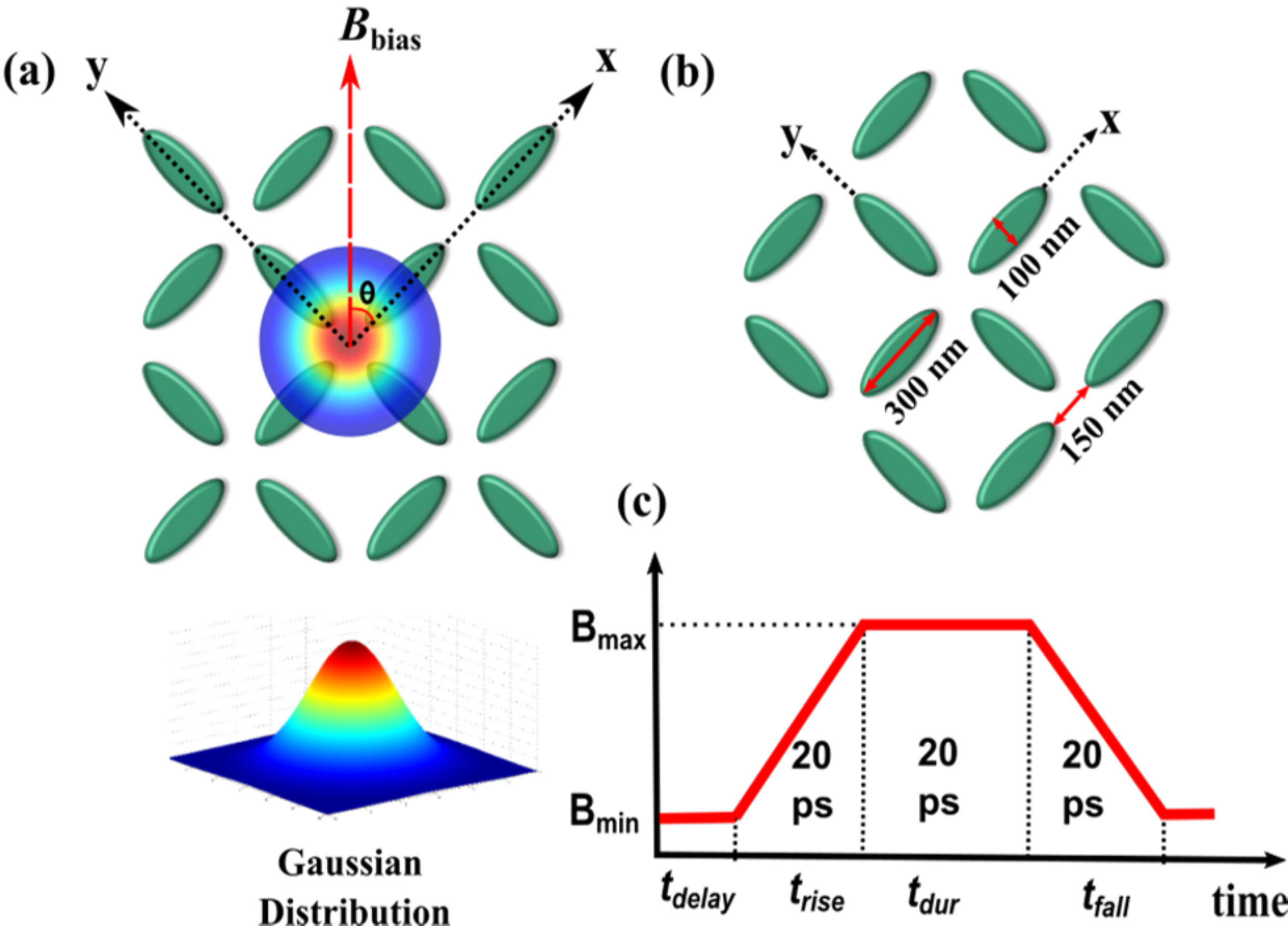}
\caption{(a) Schematic diagram of open-edged sASI system with local transient pulse excitation applied at the center vertex along the z-axis (out of the plane) and external bias field applied in the xy-plane, making an angle $\theta$ with respect to the x-axis. Inset below shows the 3D spatial profile of the local excitation, (b) Schematic diagram of the closed-edge (CE) sASI, and (c) transient square pulse with rise time ($t_{rise}$), fall time ($t_{fall}$), and duration ($t_{dur}$) of 20 ps each. nanomagnets dimensions for (a) and (b) here are: 300 nm $\times$ 100 nm $\times$ 30 nm with edge-to-edge inter-island gap of 150 nm.}
\label{fig: Intro figure}
\end{figure} 
The role of dipolar interaction in the behavior of SW propagation in sASI geometry (Fig.\,\ref{fig: Intro figure}) are investigated by performing a detailed micromagnetic analysis of propagating SWs in sASI systems with two types of vertices, viz., the ones with vertices of even coordination number while the other with vertices of mixed (even and odd) coordination number. Open-edged (OE) sASI corresponds to even coordination whereas closed-edged (CE) sASI (shown in Fig.\,\ref{fig: Intro figure}) corresponds to mixed coordination vertices. In general, the vertices with odd coordination nos. are always magnetically charged whereas the ones with even coordination may be uncharged or charged (excited state) depending on the minimum energy state corresponding to the macrospin states of the involved nanomagnets. 
\section{\label{sec:level2}Method}
For the investigations, we carried out micromagnetic simulations for the sASI structures (see Fig.\,\ref{fig: Intro figure}) using finite difference discretization based (open source) GPU-accelerated software M{\scriptsize U}M{\scriptsize AX}3~\cite{vansteenkiste2014design}. Here, the evolution of space and time dependent magnetization vector, $\vec{m}(\vec{r}, t)$, is calculated at each cell of the discretized geometry by employing Landau-Lifshitz-Gilbert equation (Eq.~\ref{eq: LLG}).
\begin{equation}\label{eq: LLG}
    \vec{\tau}_{LL} = \frac{\partial\vec{m}}{\partial t} = \gamma_{LL}\frac{1}{1+\alpha^2}[\vec{m}\times\vec{B}_{eff} + \alpha(\vec{m}\times(\vec{m}\times\vec{B}_{eff}))]
\end{equation}
with $\tau_{\text{LL}}$ is the Landau-Lifshitz torque, $\vec{m}(\vec{r}, t)$ is the reduced magnetization vector of unit length, $\gamma_{\text{LL}}$ is the gyro-magnetic ratio (rad/Ts), $\alpha$ is the dimensionless damping parameter, and $\vec{B}_{eff}$ is the effective magnetic field.\\ Here, $ \vec{B}_{eff} = \vec{B}_{ext}+\vec{B}_{demag}+\vec{B}_{exch}+\vec{B}_{anis}+ \vec{B}_{exc}(t)+...$\\
with $\vec{B}_{\text{ext}}$, $\vec{B}_{\text{demag}}$, $\vec{B}_{\text{anis}}$ and $\vec{B}_{exc}(t)$ are the external bias field, demagnetization field, anisotropy field, and time varying excitation (magnetic) field, respectively.
For the micromagnetic simulations, the nanomagnetic structures of Ni$_{80}$Fe$_{20}$ as shown in the schematic diagram Fig.\,\ref{fig: Intro figure}(a, b) are discretized in cuboidal cells of 5\,nm length, which is less than the exchange length (5.3 nm) of Ni$_{80}$Fe$_{20}$. Experimentally reported value of saturation magnetization $M_{\text{sat}} = 8.6 \times 10^{5}$\,A/m, exchange stiffness constant $A_{\text{ex}} = 1.3 \times 10^{-11}$\,J/m, and damping coefficient $\alpha = 0.5$ for $\text{Ni}_{80}\text{Fe}_{20}$ are used throughout the study~\cite{buschow2003handbook}. In order to determine the SW behavior, a reduced damping coefficient of 0.008 is used. This enables a prolonged precession of weak modes thereby allowing for a detailed analysis of such modes~\cite{barman2009dynamic}.\\
SWs are excited by perturbing the equilibrium macrospin state corresponding to a static in-plane bias field ($B_\mathrm{bias}$) via an out-of-plane temporal field pulse, see Fig.\,\ref{fig: Intro figure}. The equilibrium (macro) spins configuration is first evaluated by quasi-statically down-sweeping the external in-plane bias field from 250\,mT to $B_\mathrm{bias}$ and performing energy minimization using 4th order Runge-Kutta method. At this state, a transient square pulse field of amplitude $B_{\text{max}}= 3$\,mT with rise-time $t_{\text{rise}}$, fall-time $t_{\text{fall}}$, and pulse-duration $t_{\text{dur}}$ of 20\,ps each is applied along $z$-direction (out of the plane). The transient excitation field pulse is convoluted with Gaussian function in space to locally excite central vertex region of the OE geometry (see Fig.\,\ref{fig: Intro figure}a). After the incidence of the pulse, time evolution of reduced magnetization vector ($\vec{m}(t)$) is recorded for 4\,ns at time-step ($\Delta t$) of 10\,ps. To extract the frequencies of the excited SW modes, fast Fourier transformation (FFT) is performed on the recorded $\vec{m}(t)$ data. Thus, power spectra for SW modes are determined for a given magnetic structure. The nature and the origin of the SW modes are analyzed by investigating the power as well as phase profiles of the excited SW modes at particular frequencies using a MATLAB code (see supplementary material). For the propagational characterstics of SW, fast Fourier transform of magnetization in unperturbed neighbouring region is compared with the central perturbed vertex in two orthogonal direction. This study was performed at two external field configurations, viz., saturation and remanence for both even and mixed coordinated vertices. Additionally, to obtain a clear insight in to the role of magnetic charges on SW propagation,  a transient square pulse field is uniformly applied along the z-direction on all nanomagnets in CE as well as OE vertex geometries. SW analysis is then performed in the range of $\pm$150\,mT  of $B_\mathrm{bias}$, where $B_\mathrm{bias}$ is applied along one of the easy axes of the magnetic sublattice.\\
\begin{figure*}
    \centering
    \includegraphics[width =\linewidth]{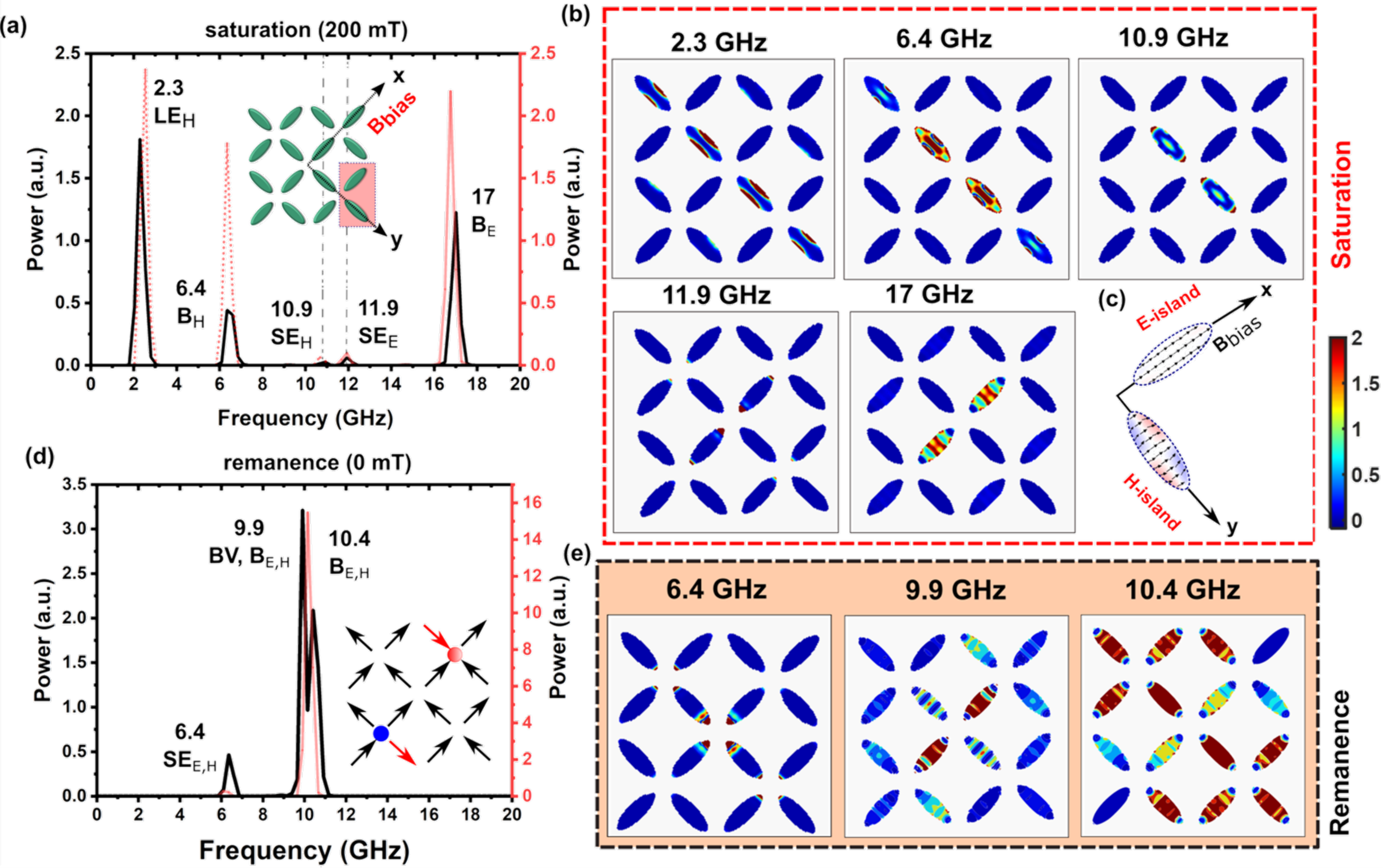}
    \caption{SW spectra (\textbf{black solid line}) for open-edge vertex  at saturation, i.e., $B_\text{bias}$ = 200\,mT applied along x-axis (a) and at remanence (d). Spectra for individual nanomagnets for bias field of 200\,mT applied along the easy axes (E-islands, \textcolor{red}{red solid line}) and the hard axes (H-islands, \textcolor{red}{red dotted line}, shown in (a))  and at remanence shown using \textcolor{red}{red solid line} in (d). Inset in (a) shows the schematic diagram for applied field direction and that in (d) shows the schematics of macro-spin configurations at remanence, (c) micromagnetic simulation results for internal spin configuration in the perpendicularly placed islands for $B_\text{bias}$ = 200\,mT as highlighted in the inset of (a). Power profile of the excited SW modes for open edge vertex at saturation (b) and remanence (e).}
    \label{fig:pos_sw_Spectra}
\end{figure*}
\section{Results and Discussion}
\begin{figure}[h]
    \centering
    \includegraphics[width = 1.0\linewidth]{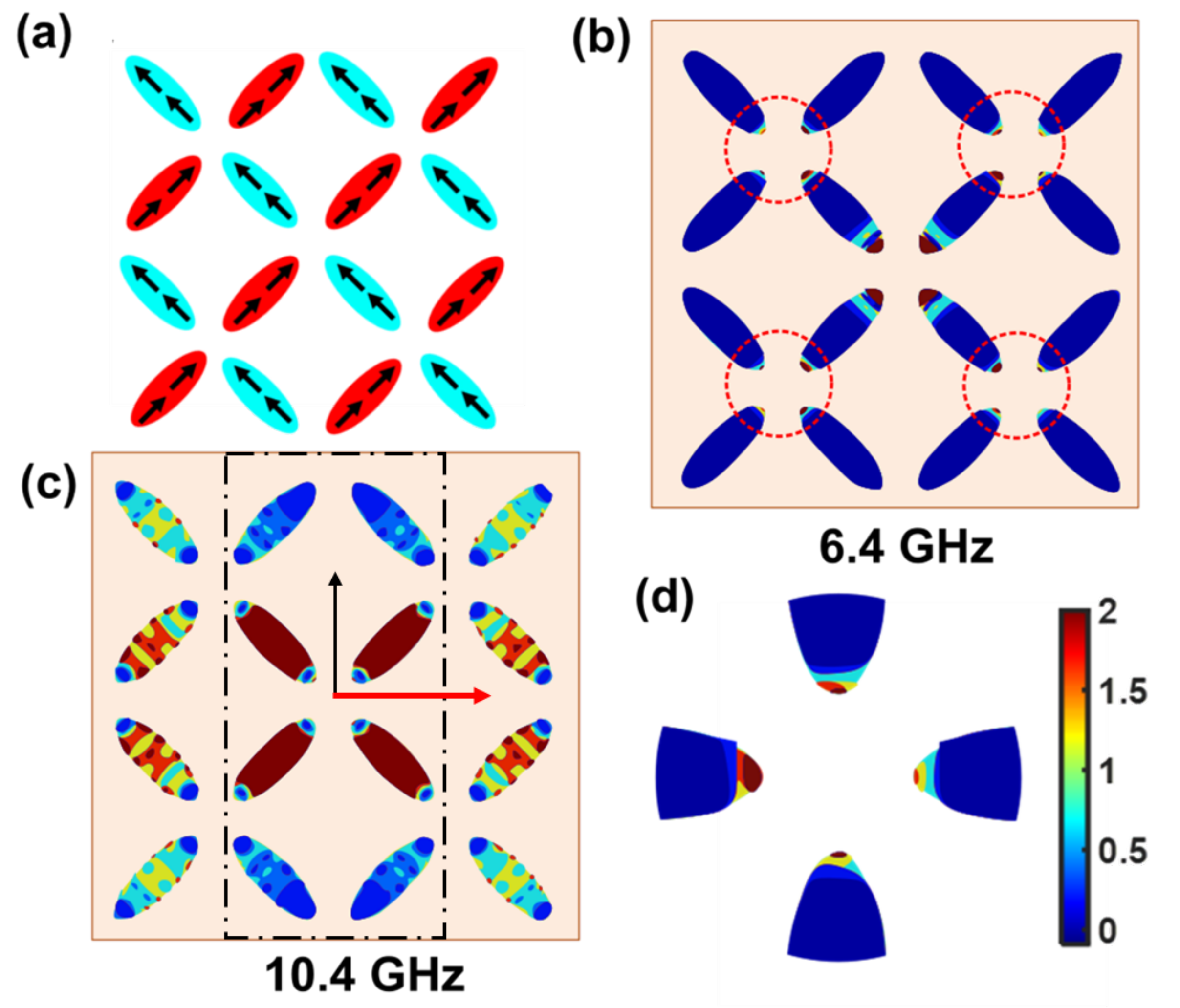}
    \caption{(a) Equilibrium magnetization state for OE vertex at remanence with $\theta=$ $45^\circ$. Power profile of the SE mode excited at 6.4\,GHz (b) and B mode excited at 10.4\,GHz (c) in remanence with $\theta=$ $45^\circ$. Black and red arrow in (c) denote the bias field orientation and its orthogonal direction. Black dash dotrectangular box encloses the nanomagnets positioned along the bias field orientation.  (d) Zoomed power profile of the vertex region highlighted with red dotted circle in (b).}
    \label{fig:angle dependence}
\end{figure}

Fig.\,\ref{fig:pos_sw_Spectra} shows the SW spectra for OE vertex geometry (vertices with even coordination) for saturated state corresponding to the bias field of $B_{\rm{bias, x}}=200$\,mT as well as remanence ($B_{\rm{bias, x}}=0$\,mT) where the bias field is applied along x-direction as clarified in Fig.\,\ref{fig:pos_sw_Spectra}c. At saturation, the spectrum exhibits five peaks corresponding to the five different modes of excitations of SWs in the frequency range of 2\,GHz\,-\,17\,GHz (see Fig.\,\ref{fig:pos_sw_Spectra}a). In the studied system involving sixteen nanomagnets, the applied field direction is along easy-axes and hard-axes of 8 nanomagnets each (see Fig.\,\ref{fig:pos_sw_Spectra}a inset), henceforth identified as E-islands and H-islands, respectively (see Fig.\,\ref{fig:pos_sw_Spectra}c). As the orientation of the external bias field applied in this case corresponds to easy as well as hard directions of different nanomagnets, we have analyzed SW behavior of \textit{individual} islands with external field along both easy and hard directions . This allows us to carry out an in-depth analysis of the role of individual nanomagnets in the overall SW spectra of the dipolar coupled system. The results are shown in Fig.\,\ref{fig:pos_sw_Spectra}(a). A comparison of the SW modes excited in the individual nanomagnets with that of the coupled sASI system (Fig.\,\ref{fig:pos_sw_Spectra}(a)) shows that although the modes associated with coupled system display a nominal frequency shift of up to $\sim$\,0.3\,GHz which accounts for the stray field of $\sim$\,3.6\,mT, yet it consists of all the modes excited in individual E-as well as H-islands. The shift may be a result of the influence of additional dipolar field emanating from the interacting nanomagnets. Thus, these results suggest that as the spins are strongly pinned along the direction of the $B_{\rm{bias}}$ at saturated state, the SW modes in individual nanomagnets remain largely unaffected in the dipolar coupled sASI system.\\ 
In order to determine the origin of these peaks in the observed power spectra, we analyze the power profiles of the excited SW modes corresponding to each of the five frequencies where a peak is observed. The power profiles are shown in Fig.\,\ref{fig:pos_sw_Spectra}b.
We make the following four interesting observations: i) significant mode-power for the three lower-frequency modes viz., 2.3\,GHz, 6.3\,GHz, and 10.9\,GHz, appears to be concentrated primarily in the H-islands. This indicates that the peaks observed at those frequencies are primarily due to modes excited in the H-islands. ii) with increasing frequencies, inhomogeneously distributed power is systematically reduced in the H-islands. For the SW mode at 2.3\,GHz, excited mode exhibits significant power in the four H-islands, whereas for the mode at 10.9\,GHz, the power is found to be mainly concentrated in the two H-islands coupled to the central vertex. iii) two higher frequency modes appearing at 11.9\,GHz and 17\,GHz are primarily excited at the E-islands. Large power corresponding to these modes is primarily located at the E-islands of the central vertex. iv) the excited SW modes in both E- and H-islands are inhomogeneously distributed. In general, demagnetization field in confined magnetic nanostructures may be inhomogeneous which results in multiple low energy potential wells where spins are in local equilibrium corresponding to the internal local field. Curling of magnetization observed near the edges of H-islands is an example of one such inhomogeneity of demagnetization field (see Fig.\,\ref{fig:pos_sw_Spectra}c). Such inhomogeneous demagnetizing field within the individual sASI nanostrctures may in turn lead to inhomogeneously distributed local excitation of SWs in the sASI structures. According to the spatial distribution of the modes, we identify the modes observed at edges of nanomagnets as edge-modes and the others excited within the ``bulk" of the nano-magnets as bulk-modes (hereafter B-mode). Thus, we identify the 2.3\,GHz mode as long-edge ($\rm{LE_H}$) mode and 11.9\,GHz as short-edge ($\rm{SE_E}$) mode. The suffices denote the H- and E-islands, respectively.\\
From the power distribution profile at saturation we also observe that the lower-frequency $\rm{LE_H}$ modes appear to extend to the nearest neighbors (H-islands) restricted along the direction perpendicular to that of $B_{\rm{bias}}$. The demagnetization field for H-islands is larger than that for the E-islands thereby reducing the effective in-plane field ($B_{\rm{eff}}$) in H-islands compared to that for E-islands. The total field $\vec{B}_{\rm{tot}} (=\vec{B}_{\rm{eff}}+\vec{B}_{\rm{exc}}(t)$) during temporal pulse excitation along z-direction thus make a larger angle with respect to the in-plane direction of $B_{\rm{eff}}$ resulting in larger fluctuations in z-component of magnetization in H-islands. Since $f\,\propto B_\mathrm{tot}$, therefore, the perturbation in phase coherent precession which manifests itself as a SW has higher probability of transmittance/propagation to the nearest neighbour of H-islands which in this case along the y-axis for lower frequency mode $\sim 2.3$\,GHz and 6.4\,GHz. Along the x-axis, adjacent H-islands are separated by much larger distance $\sim 450$\,nm. At this separation, resulting stray field variation due to fluctuations in z-magnetization is negligible. Thus, we don't observe transmittance of SW mode along field direction or x-axis.\\
At remanence ($B_{\rm{ext}}=0$), the magnetization in each 
nanomagnet orients along their easy directions. Due to the magnetostatic
interactions, low-energy 2-in/2-out spin ice states are observed for 
three vertices whereas the remaining two vertices exhibit magnetically 
charged excited states, see inset of Fig.\,2(d). The magnetic charges are reconciled 
considering the nanomagnets as a dumbell of two equal and opposite 
magnetic charges (dumbell model)~\cite{castelnovo2008magnetic}. For this magnetic configuration of sASI system at remanence, number of SW modes are now reduced to three from five as compared to the saturated state. In this case, both E- and H-islands are now identical with respect to an external transient field perturbation which accounts for the observed less nos. of peaks in remanence. The peaks 
are observed at the frequencies of 6.4\,GHz, 9.9\,GHz, and 10.4\,GHz, 
respectively in the SW-spectra as shown in 
Fig.\,\ref{fig:pos_sw_Spectra}d. In order to clearly analyze the role of dipolar interactions in the SW spectra in this case, we have calculated the SW modes for \textit{individual (i.e., non interacting)} nanomagnets with magnetization along the easy directions. The simulation results show that only two SW modes, viz., 6.4\,GHz and 10.4\,GHz, are excited in an individual nanomagnet of same dimensions thereby indicating a possible role of dipolar interactions on the excitation of 9.9\,GHz-mode in such dipolar coupled nanostructures.
\begin{figure}[ht]
    \centering
    \includegraphics[width = 1.0\linewidth]{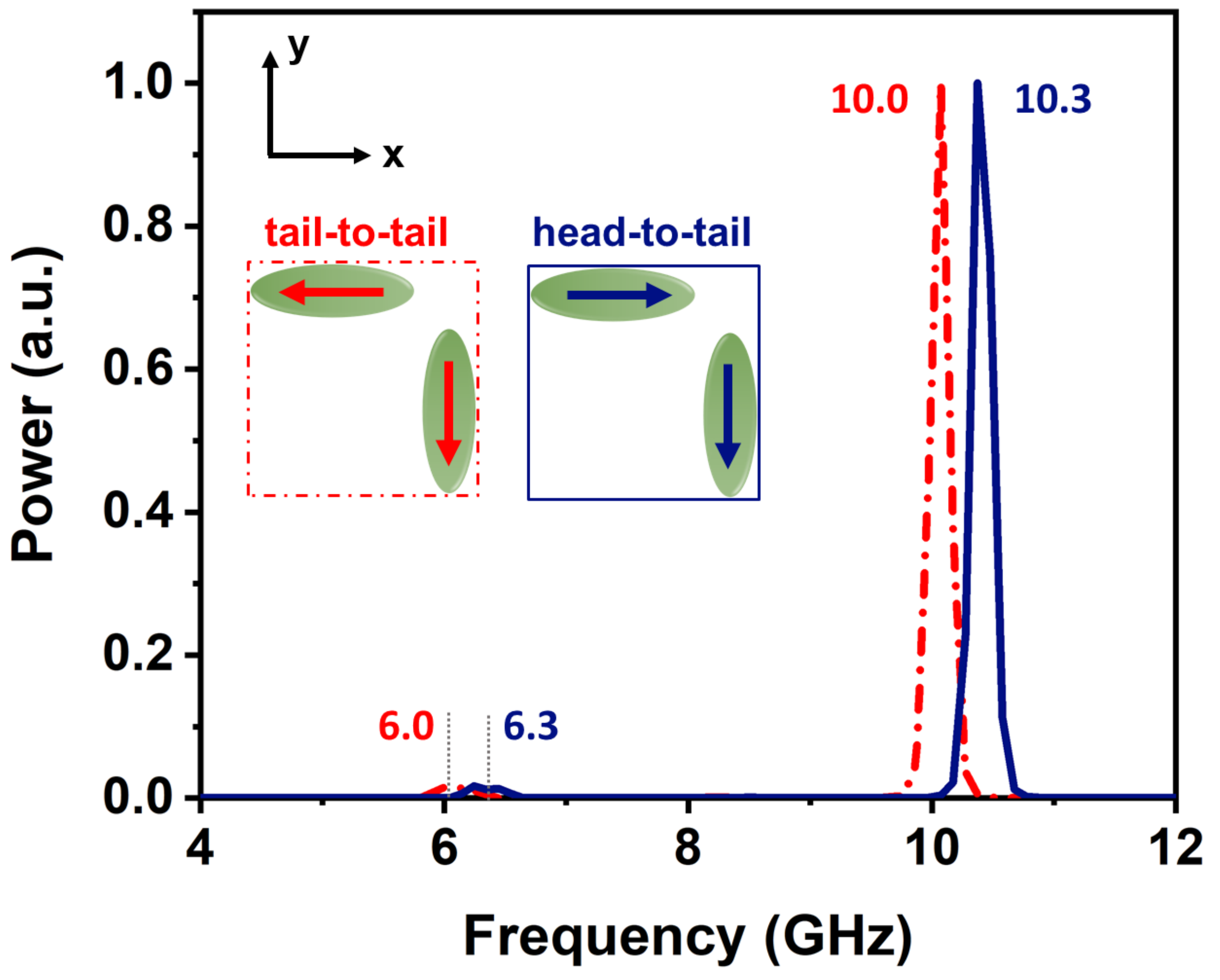}
    \caption{SW spectra of L-shaped geometry at remanence in two magnetic state configuration viz., tail-to-tail (magnetically charged vertex) and head-to-tail (magnetically uncharged vertex). \textit{Inset} shows the xy orientation for the opted geometry and magnetic state with tail-to-tail as well as head-to-tail configuration.}
    \label{fig:L-shape SW}
\end{figure}
Therefore to elucidate the origin of this 9.9\,GHz mode, we analyze a fundamental cell consisting of two dipolar-coupled nearest neighbour nanomagnets: one E- and one H-island in L-shaped geometry (see Fig.\,\ref{fig:L-shape SW}). The SW spectra of the L-shaped geometry at remanence, where the magnetization in the islands orient as head-to-tail configuration resulting in zero magnetic charge at the meeting point (junction) of the nanomagnets, exhibit two peaks at the frequencies of 6.3\,GHz and 10.3\,GHz, respectively. Since the individual nanomagnets show similar SW behavior, these observations suggest that such head-to-tail (i.e., zero magnetic charge) configuration in L-shape geometry retains the SW behavior of individual nanomagnets. However, as the magnetization of the L-shaped geometry is predefined as head-to-head or tail-to-tail configurations leading to a magnetic charge at the junction, the SW spectra show a red-shift of $\sim 0.3$\,GHz with respect to that observed for individual nanomagnets thereby exhibting a strong mode at 10\,GHz. This demonstrates a clear role of magnetic charges due to dipolarly interacting nanomagnets on the excitation of SWs. 
This 10\,GHz-mode appears to be excited at 9.9\,GHz in the sASI system with mixed types of charged and chargeless vertices. This suggests that SW spectra for vertices with mixed charge state may exhibit SW modes as observed in Fig.\,\ref{fig:pos_sw_Spectra}d. In order to elucidate this observation further, simulations were carried out for the five-vertex sASI system containing predefined chargeless vertices. We observe an absence of the additional mode at 9.9\,GHz in the SW spectra (see supplementary materials, Fig. S1) which further demonstrates that the origin of the 9.9\,GHz peak in the SW spectra lies in the magnetically charged excited vertices. These results also underlines the fact that the magnetostatic interaction between nanomagnets leading to 3-in/1-out or 3-out/1-in type local charged vertices modifies the local magnetic structure of the involved nanomagnets in such a way that an additional SW mode is excited in these nanostructures. In the following, we identify this mode as charged vertex (CV) mode.\\
Further confirmation on the role of strong dipolar interaction in the excitation of the additional SW mode is obtained from the investigation of the SW modes of the OE vertex system with varying inter-island separations. The spectral behavior was investigated by varying the edge-edge separation of the co-linear nanomagnets from 100\,nm to 1\,$\mu$m. We observe that the static magnetic behavior (i.e., macrospin configuration) of the sASI system remains unaltered in the entire range of separation (i.e., 100\,nm-1$\mu$m). However, the excitation of the additional mode at 9.9\,GHz is observed till the separation of $\sim225$\,nm (see supplementary information for details) beyond which only the two modes, as excited for individual nanomagnets, are observed. Based on these observations, we identify the separation up to 225\,nm as the strongly interacting regime. Next, in order to investigate the nature of the excited SW modes at remanence in details, the power and phase profiles of the modes are calculated. As shown in Fig.\,\ref{fig:pos_sw_Spectra}(e), excitations for the three frequencies are observed in both type of nanomagnets. However, the modes can be distinctly identified according to the distribution of the power profiles within the individual nanomagnets. The excitation of 6.4\,GHz mode appears to be localized near the short edges (SE mode) whereas the 9.9\,GHz and 10.4\,GHz modes are excited in the bulk of the nanomagnets. The power profiles clearly show that the modes which are excited in the nanomagnets at the central vertex, propagate beyond the central vertex albeit with reduced intensity and different propagational characteristics (see below).\\
\begin{figure}[h]
    \centering
    \includegraphics[width = 1.0\linewidth]{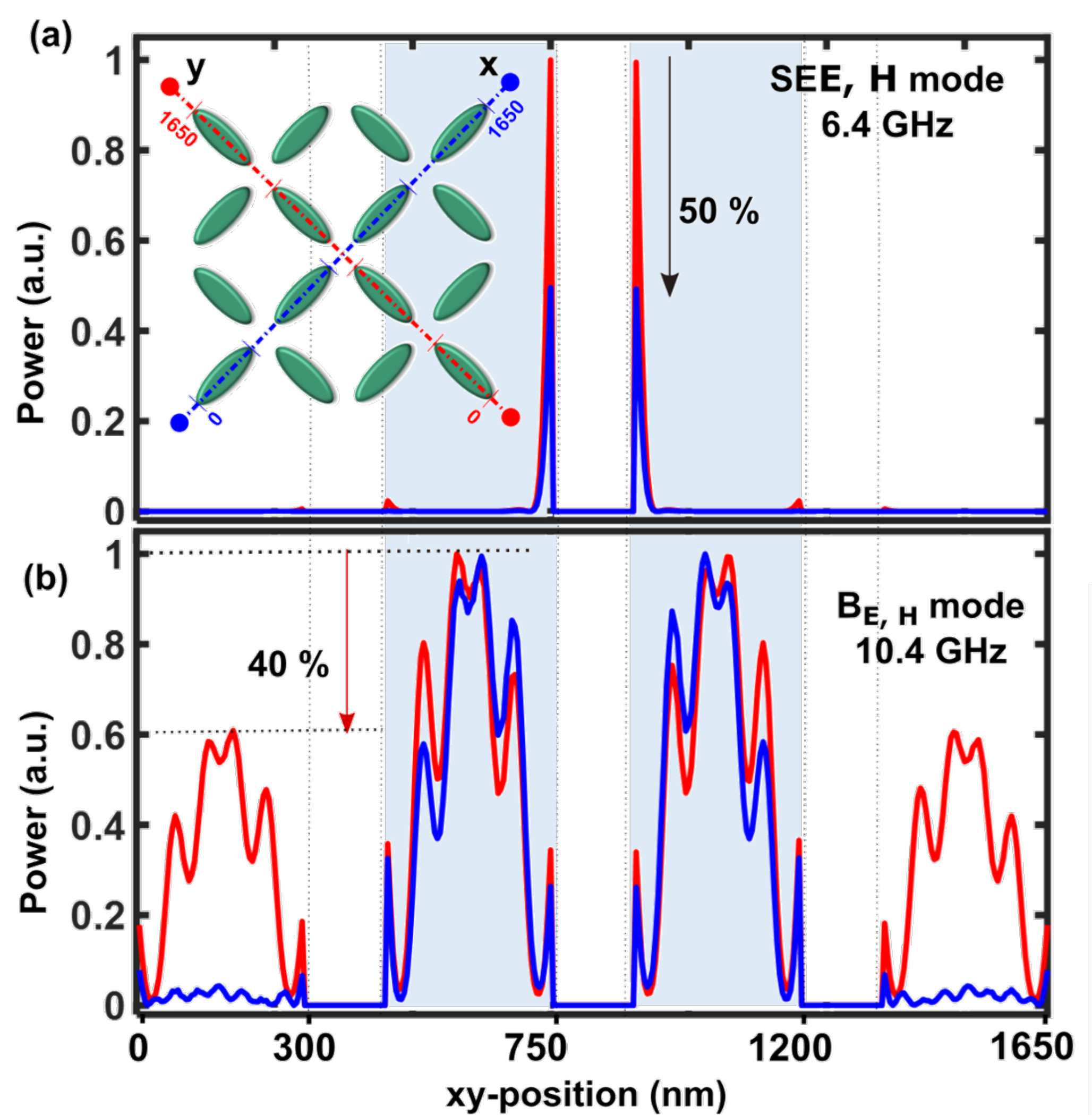}
    \caption{Line power profile for the excited SW mode at (a) 6.4\,GHz and (b) 10.4\,GHz in remanent state, along the two orthogonal directions as shown in the inset of (a) (\textcolor{blue}{Blue}: along x, \textcolor{red}{Red}: along y). Highlighted regions correspond to nanomagnets at central vertex which are under pulse field excitation.}
    \label{fig:line profile dipole sw}
\end{figure}
The SE 6.4\,GHz mode exhibits stronger power along the direction perpendicular to the bias field than along the field direction. This is exemplified by plotting the line profile of the power map (Fig.\,\ref{fig:pos_sw_Spectra}e) along the x- and y-directions through the central vertex as shown in Fig.\,\ref{fig:line profile dipole sw}. The line profile clearly shows that the power of the 6.4\,GHz mode excited at the central vertex nanomagnets along the x-direction is reduced by almost 50\% as compared to that along y-direction. A nominal power of $\sim2$\% at the other end of these nanomagnets is observed only along y-direction. No power for excitation is observed in the nanomagnets of outer vertices. Thus, the excited SE mode is strongly localized at the central vertex where the pulse field is applied and doesn't exhibit propagational character in dipolar-coupled nanostructures. For the bulk SW mode of $\sim10.4$\,GHz, the observed power of excitation at the two nanomagnets involved at the central vertex is nearly the same for both E- and H-islands. However, for nanomagnets involved at the outer vertices, the observed power along y-axis is nearly 60\% of the maximum power observed at central vertex (see Fig.\,\ref{fig:line profile dipole sw}b). whereas a nominal power of only $\sim2$\% of the maximum power is observed at the nanomagnets of the outer vertices along the field direction.. Thus, we observe a significant SW propagation of the bulk 10.4\,GHz mode along the direction perpendicular to the applied bias field direction (i.e., x-direction). For the CV mode observed at 9.9\,GHz, the power profile is significantly different than the other two modes. Here, the power is found to be concentrated mainly at the nanomagnets involved at the central vertex and the nanomagnets with reversed magnetization (-y direction) at the outer vertices with non-zero local charges (see Fig.\,\ref{fig:pos_sw_Spectra}e and inset of Fig.\,\ref{fig:pos_sw_Spectra}d). Here, the power profile of this mode in the H-islands of the central vertex exhibit nodal lines perpendicular to the magnetization orientation. This profile is the characteristic of backward volume like mode (BV). However, power of the mode is distributed in the bulk of the rest of the nanomagnets. Thus, 9.9\, GHz mode can be identified as hybridized BV and B mode.   \\ The above analyses suggest that the SW modes (SE- and B-modes) in such unconnected highly shape anisotropic nanomagnets in sASI geometry show preferential propagation along the direction perpendicular to the external bias field direction. We note here that at ground state, this is also the direction along which the two charged vertices are present in the system. Thus, so far it is not clear if the propagation direction is affected by the external field or the presence of magnetic charges. In order to elucidate this further, we investigated the SW behavior of the sASI system as a function of applied field direction. By applying external bias field along different directions, the remanent magnetic states of the individual nanomagnets are tuned. The SW modes for these investigations are calculated for $\theta =45^\circ, 90^\circ,$ and $135^\circ$, respectively where, $\theta$ is the angle between the applied field direction and x-axis (see Fig.\,\ref{fig: Intro figure}a). 
For $\theta =45^\circ$, the magnetic configuration of the sASI system at remanence exhibits chargeless 2-in/2-out vertices in the sASI system as shown in Fig.\,\ref{fig:angle dependence}a. The SW spectra for this chargeless sASI state exhibit only two modes which are excited at 6.4\,GHz (SE mode) and 10.4\,GHz (B-mode), respectively. Interestingly, excitation of only these two modes were earlier observed for chargeless L-shaped geometry as discussed above. Here we observe a remarkable difference in the power profile for 6.4\,GHz mode calculated for this magnetic configuration corresponding to $\theta =45^\circ$ in comparison to that observed for $\theta =0^\circ$. As shown in Fig.\,\ref{fig:angle dependence}b,d, the SE mode for this magnetic configuration is found to be excited in all the nanomagnets whereas for $\theta =0^\circ$ the mode-power at the two farthest nanomagnets along x-axis is nearly zero suggesting no excitation of the SW mode in those islands.
\begin{figure*}
    \centering
    \includegraphics[width = 0.85\linewidth]{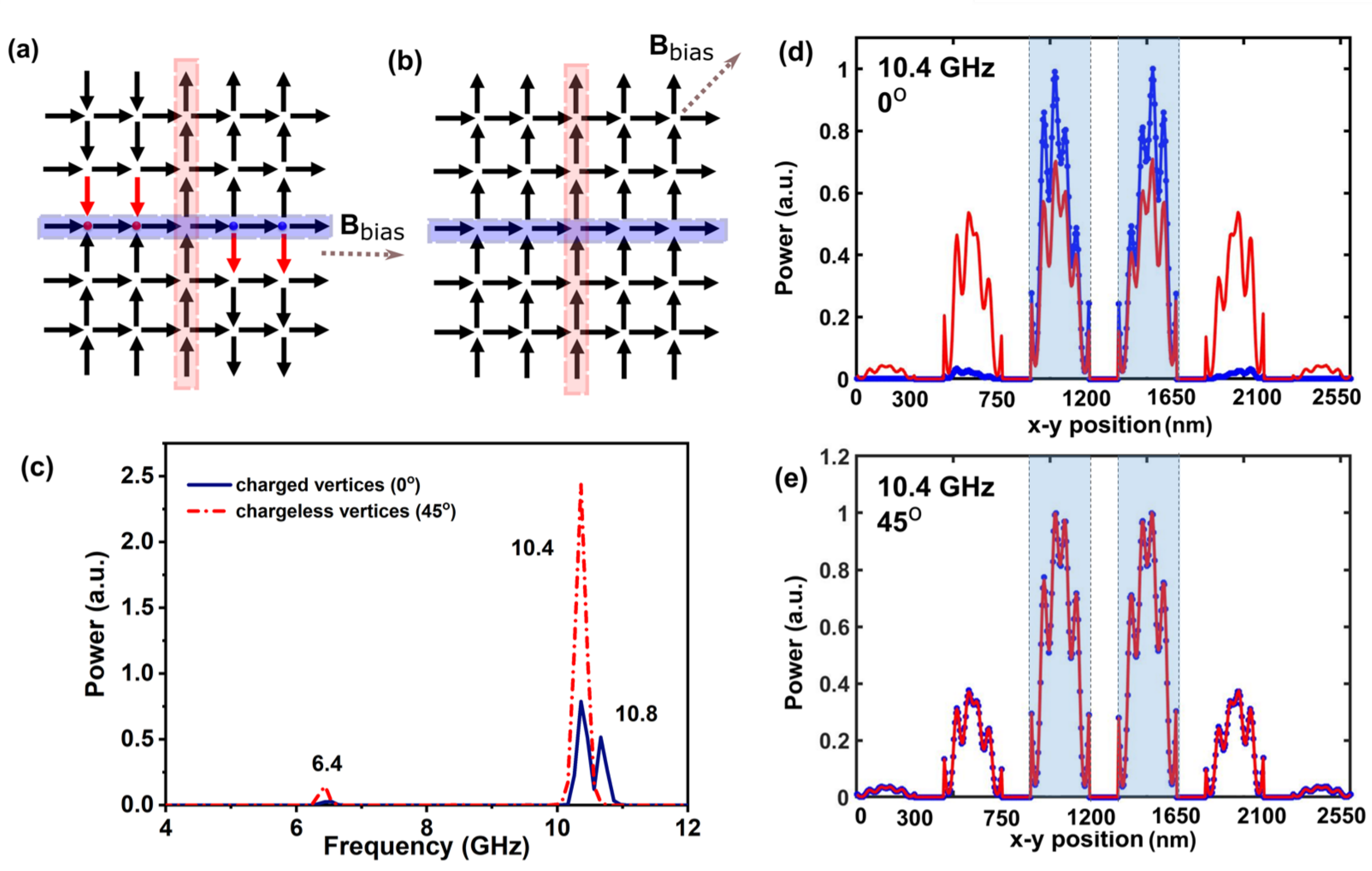}
    \caption{Equilibrium macrospins configuration in sASI array at remanence when initial bias field applied along one of the easy axes , i.e. 0$^\circ$ (a) and at 45$^\circ$ (b) with respect to the easy axis of one of the sublattices, (c) SW spectra for 5$\times$5 vertex array for equilibrium state shown in (a, b). Line power profile of the bulk SW mode excited in sASI array at remanence, across the two orthogonal direction (highlighted region shown in (a, b)) when initial bias field is oriented along one of the easy axes, i.e. 0$^\circ$ (d) and at 45$^\circ$ (e) from the easy axis of one of the sublattice. Highlighted regions in (d, e) correspond to central vertex nanomagnets which are under field pulse excitation.}
    \label{fig:array_dipole}
\end{figure*}
For this chargeless state at $\theta=45^\circ$, the sASI system is magnetically symmetric with respect to the applied bias field. Therefore, the SE mode is excited symmetrically in all the nanomagnets. However, for $\theta=0^\circ$, the sASI system is magnetically asymmetric with respect to the bias field. Under this condition, magnetically charged sASI vertices are generated as discussed above. Our analysis suggests that these charged vertices for $\theta=0^\circ$ are responsible for the observed asymmetric propagation of the SE mode. For 10.4\,GHz B mode, it is apparent (see Fig.\ref{fig:angle dependence}c) that the nanomagnets of the outer vertices positioned along the field direction (enclosed in black dash-dot rectangular box) display nominal power as compare to the nanomagnets positioned perpendicular to the field orientation. This observation is similar to the case of $\theta = 0^{\circ}$ where 10.4\,GHz mode propagates along the orthogonal to the bias field orientation. Thus, the 10.4\,GHz B-mode exhibits an asymmetric propagation with respect to bias field.  
Absence of the CV mode in this case of $\theta=45^{\circ}$ clearly validates our observations for a significant role of dipolar interactions on the SW excitation in this system. 
Our additional calculations performed for $\theta =135^\circ$ shows a similar behavior of the SW modes as for $\theta=45^\circ$. For $\theta =90^\circ$ i.e., when the bias field is applied along y-direction (see supplementary Fig.S3a), the remanent state exhibits two charged (outer) vertices positioned along the field direction. Thus, these charged vertices for $\theta =90^\circ$ appear in orthogonal direction to the charged vertices observed for $\theta =0^\circ$ case. Interestingly, the SW spectra for this case again show the CV mode in addition to the SE and B-modes (see supplementary information). Thus, for this charged state also, three modes are excited as for $\theta =0^\circ$ with similar SW propagational characteristics. These results clearly demonstrate the following three facts: one, the excited SWs in such unconnected nanomagnets propagate to other nanomagnets which are within the strongly interacting (dipole-dipole interaction) regime which is $\sim225$\,nm of interisland separation. Remarkably, this suggests that the dipolar coupling of the nanomagnets facilitates carrying the specific information of SWs from one nanomagnet to another. Secondly, the presence of ``sources" and ``sinks" of magnetic flux favors excitation of a specific mode - 9.9\,GHz for the given structure. Third, the propagation of B-mode (10.4\,GHz) is tuned by the orientation of bias field. Also, we observe that the propagation of SE mode is impeded by the ``source" or ``sink" of magnetic flux (see Fig.\,2(e)). Thus, our results show that by tuning charged magnetic state in such structures along with bias field orientation, guided or directional SW propagation can be realized  in such unconnected nanostructures.\\
Although the above analysis shows SW propagation in the unconnected nanostructures, the distance up to which SW propagates is not clear. In order to investigate this, we have calculated the behavior of SWs for a larger array of vertices. An array of 5$\times$5 vertices at remanent state with two different initial bias field configurations is considered for the analysis. For $\vec{B}_{\rm{bias}}$ applied along one of the easy axes of the nanomagnets, the magnetic configuration at remanent state exhibits two pairs of charged vertices (i.e., two ``sources" and two ``sinks" of flux) separated by an uncharged 2-in/2-out type vertex along the external field direction as schematically depicted in Fig.\,\ref{fig:array_dipole}a. Thus, we observe similar micromagnetic behavior as for the 5-vertex case discussed above except that in this case two pairs of charged vertices instead of one pair are observed for this array of 25 vertices. To observe the SW propagation, again we apply a square pulse field with spatial Gaussian distribution of similar specification as before at the central uncharged vertex (see Fig.\,\ref{fig:array_dipole}a) and calculate the SW spectra for the system. Here again, the calculated SW spectra exhibit three excited modes, however, at frequencies of 6.4\,GHz (SE-mode), 10.4\,GHz (CV-mode), and 10.8\,GHz  (B-mode) as shown in Fig.\,\ref{fig:array_dipole}c. The power profiles are shown in the supplementary material in Fig. S4. In this case, 9.9\,GHz and 10.4\,GHz mode as discussed above for 5-vertex state shift to 10.4\,GHz and 10.8\,GHz, respectively most likely due to the different net stray field emanating from the constituting islands. For $\vec{B}_{\rm{bias}}$ applied along 45$^\circ$, the magnetic configuration at remanence exhibit all 2-in/2-out chargeless vertices (see Fig.\,\ref{fig:array_dipole}b) as in the case of 5 vertices discussed earlier. SW spectra for this state again exhibit two modes at 6.4\,GHz and 10.4\,GHz similar to 5 vertices case which corresponds to SE, and B-mode (see supplementary info for power profiles). Thus, the calculated SW spectra for an array of 25 vertices corresponding to two different magnetic configurations stabilized for $B_{\rm{bias}}$ applied at 0$^\circ$ and 45$^\circ$ further confirm the important role of charged vertices in the origin of the additional peak corresponding to the CV mode.\\
In order to obtain more quantitative information about the propagational characteristics of SW modes in an array of 25 vertices, we investigate the power profiles (see supplementary material) of the modes across the two orthogonal directions highlighted in Figs.\,\ref{fig:array_dipole}a,b. Line profile of power for magnetic configuration for $\theta=0^{\circ}$ (array with charged vertices) suggests that  a significant SW power propagate along the y-direction 
whereas for $\theta= 45^\circ$, SWs propagate symmetrically in both the x- and y-direction (see Fig.\,\ref{fig:array_dipole}d,e) owing to the geometrical symmetry of the system. To quantify the amount of propagation to the nearest vertex, we calculate the transmission coefficient ($\rm{P_{neighbor-island}}/P_{\rm{centre-island}}$) in both the aforementioned cases where $P_{\rm{centre-island}}$ and $P_{\rm{neighbhor-island}}$ are corresponding powers at central and a neighboring island, respectively. For $\theta=0^{\circ}$, we observe that the B-mode shows efficient transmission to the nearest neighbour vertices with transmission coefficient $\sim$\,0.75 towards the direction perpendicular to the initial bias field which reduces to 0.07 to the next nearest-neighbour vertex. However, along the field direction or direction of charged vertices,  transmission coefficient of only $\sim0.02$ is observed indicating insignificant transmission along the direction of charged vertices. 
For $\theta=45^{\circ}$ with no charged vertices, symmetric transmission along both the orthogonal directions with transmission coefficient  of $\sim$0.4 at nearest neighbor (see Fig.\,\ref{fig:array_dipole}e) is observed owing to the geometrical symmetry of the system. However, farthest nanomagnets of the nearest vertices positioned perpendicular to the field direction, display no power as compared to the nanomagnets positioned along the field direction (see Fig.S4d of the supplementary info). These results confirm that although the bias field orientation plays a role in directional propagation of SWs, presence of charged vertices along with it significantly enhance the propagated power to the nearest vertex. Also a profound SW transmission of $\sim75\%$ can be observed upto only first nearest neighbour thus limiting the transmission upto $\sim450$ nm (center-to-center vertex distance) in our case. The SE mode, however, behave similarly as for OE vertices discussed before (see Fig.\,S4c,d of supplementary info). For $\theta=0^\circ$, propagation of SE mode is impeded along the charged vertices whereas a symmetric propagation of SE mode is present for chargeless vertices at $\theta=45^\circ$.
\begin{figure*}
    \centering
    \includegraphics[width = 0.85\linewidth]{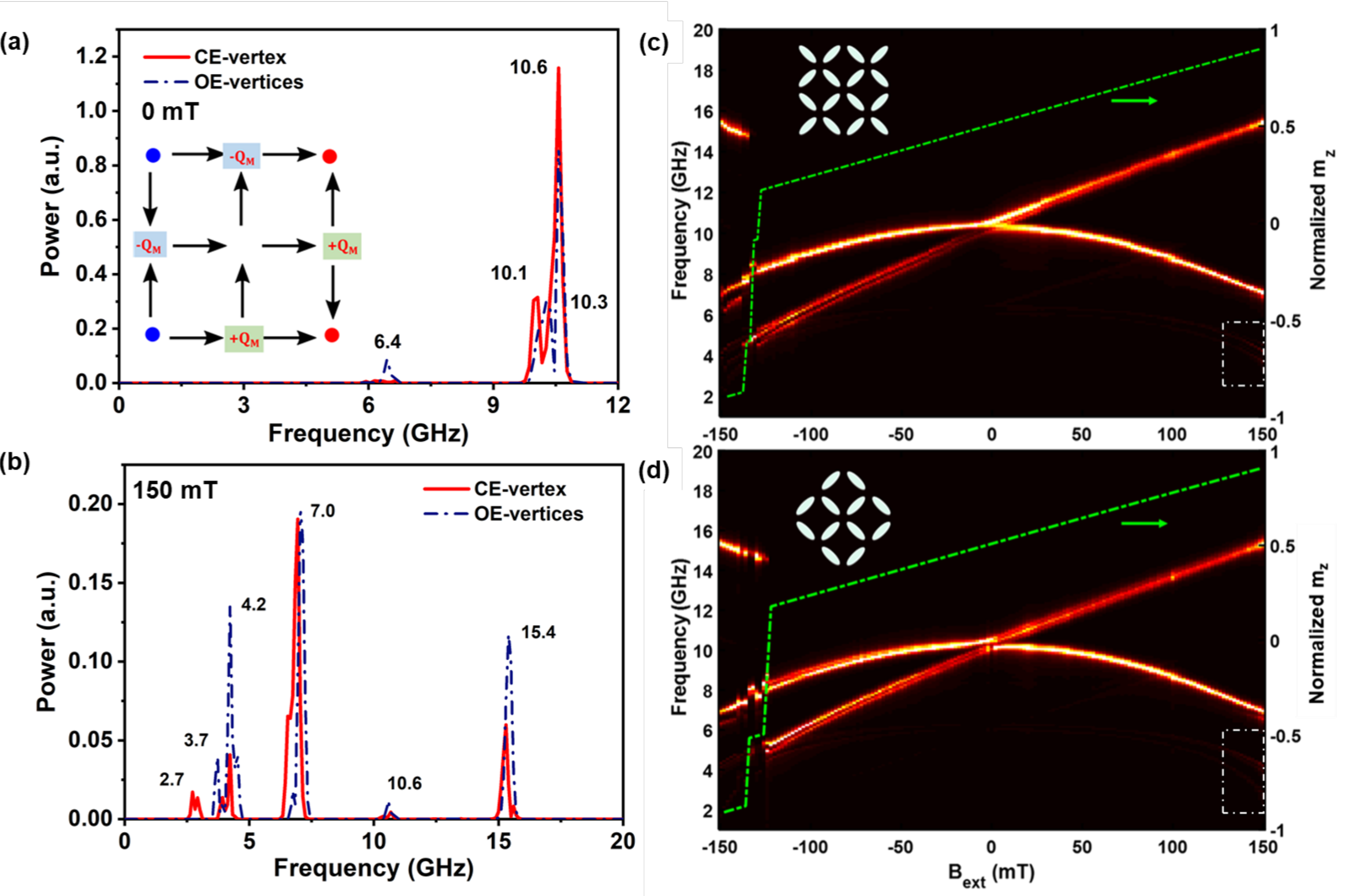}
    \caption{SW spectra for CE and OE vertices at remanence, i.e., $B_{\rm{bias}}$=0\,mT (a) and
at $B_{\rm{bias}}$=150\,mT (b) applied along one of the easy axes. Inset in (a) shows the schematic of macrospins arrangement and uncompensated charges at odd coordination vertices for CE vertex at remanence. (c) Dependence of SW mode frequency on bias field strength applied along one of the easy axes (left ordinate) and normalized magnetization vs bias field strength (right ordinate) plot during down-sweep for OE-vertices and (d) CE-vertex. white dash dot rectangular box show the position of weaker modes which are not apparent in the surface plot.}
    \label{fig:closed edge magnetic state}
\end{figure*}
\\In large arrays of such sASI systems, often the structures are fabricated such that additional islands are patterned at the edges, thus forming CE systems, see Fig.\,\ref{fig: Intro figure}(b) for a schematic diagram. To complete our understanding of SWs in sASI systems, we next investigate the SW excitation in CE-vertex. We note here that unlike for OE-vertices as discussed above, here the vertices are of mixed coordination. The vertices at edges are of coordination no. 3 and that at the centre is of 4 (see Fig.\ref{fig: Intro figure}(b)). This results in an uncompensated magnetic charge at the odd coordinated 1-in/2-out or 2-in/1-out type vertices. Our earlier work~\cite{keswani2020accessing} shows that for such mixed coordinated vertices, the net distribution of charges at outer vertices screen and compensate any charge that is formed at inner/central even coordinated vertex. The cloud of charges at the border then form  magnetic polarons~\cite{keswani2020accessing, chern2016magnetic, gilbert2014emergent} which is otherwise absent in case of OE vertices. Therefore, the behavior of SWs in magnetopolaronic environment is an open question which we address here.\\ 
Accordingly, we investigate the feasibility of forming magnetic-polaron and its influence on SW dynamics during an external bias field sweep within $\pm$150\,mT. The field range of $\pm$150\,mT is chosen such that underlying spins are in non-saturated state and therefore the role of dipolar interactions is significant in stabilizing the magnetic state at each bias field. Our simulation results for CE vertex show that within the aforementioned bias field regime, CE vertex does not exhibit non-zero magnetic charge at the center vertex whereas there is still a magnetic charge cloud comprising of uncompensated $\pm Q_m$ charges (see inset of Fig.\,\ref{fig:closed edge magnetic state}a) such that net magnetic charge of the system remains zero. The SW spectrum for this magnetic state at remanence shows two SW modes at 10.1\,GHz and 10.6\,GHz. However, spectrum for OE vertices under uniform excitation at remanence shows three SW modes at 6.4\,GHz, 10.3\,GHz, and 10.6\,GHz. Interestingly, we observe that weak mode at 6.4\,GHz for CE vertex with charge cloud is absent and 10.3\,GHz mode is red-shifted to 10.1\,GHz. The red-shift may be accounted due to the variation in the dipolar field environment for charge cloud in CE-vertex and localized charge in OE-vertices. The absence of 6.4\,GHz mode further uphold our earlier statement that presence of local magnetic charges may impede the propagation and excitation of SE mode (6.4\,GHz). Details of the mode-profiles are given in supplementary information. We next determined the evolution of the SW spectra in varying field.  For comparison, the resulting SW spectra for the two types of vertices viz., the CE and the OE vertices (discussed above) at varying fields are shown in Fig.\,\ref{fig:closed edge magnetic state} as 2D surface plot. The Figs.\,\ref{fig:closed edge magnetic state}(c) and (d) represent the evolution of SW modes, i.e., their excitation frequencies as a function of static bias field applied along one of the easy axes of sASI. The simulations were carried out at every 2\,mT field in the entire field range of $\pm$150\,mT. Initially at 150\,mT, both CE and OE vertices display two strong modes at $\sim$7.0\,GHz and 15.4\,GHz and three weak modes at $\sim$ 3\,GHz, 4.2\,GHz, and 10.6\,GHz, respectively. Weak modes are not distinctively visible in 2D surface plot due to very low relative intensity. The position of these modes are shown by a rectangular dash dot box. For clarity, SW spectrum of CE and OE vertices at 150\,mT is shown in Fig.\,\ref{fig:closed edge magnetic state}(b). From the individual spectrum at 150\,mT, weak modes can be easily identified. Furthermore, we observe that there is a slight shift in the frequencies ($\sim$0.1\,GHz) of the observed sw modes which become significant($\sim$1\,GHz) for the lower frequency mode at 3.7\,GHz. Thus, mixed and even coordinated vertices (CE and OE vertices) display nominal discrepancy in the frequencies of the SW mode which may become significant for the lower frequency mode. Details of the power profiles are provided in supplementary information.   
Further, we focus on the behavior of two prominent SW modes appearing at 15.5\,GHz and 7\,GHz which exist during complete field sweep regime and exhibit different field dependence. Here, 15.5\,GHz mode is the fundamental bulk mode associated to E-islands whereas 7\,GHz mode is the fundamental bulk mode associated to H-islands. Bulk mode associated to E-islands shows a linear response from 150\,mT to a negative bias field strength of -122\,mT and -128 mT for CE and OE vertices, respectively, with a positive slope during down-sweep. The observed SW response on external field sweep can be understood by examining the variation in internal spin configuration in E-islands. During the investigated field sweep regime, magnetic spins remain aligned along the easy-axis and display a first sharp switching at -128\,mT (OE vertex) and -122\,mT (CE vertex) when Zeeman field prevails over anisotropy field (see magnetization vs field plot superimposed on 2D SW frequencies vs field surface plot in Fig.\,\ref{fig:closed edge magnetic state}).
Thus, E-islands are subjected to decreasing effective field during down-sweep of external bias field with underlying spin configuration remains invariant. Hence, frequency of fundamental SW mode arisen due to uniform spin alignment follow linear trend in accord with larmour precession $f\propto H_{\rm{eff}}$ up-to switching field. As the switching appears, fundamental SW mode corresponding to E-island appearing at $\sim$5\,GHz shifted to $\sim$14\,GHz due to sudden variation in underlying macrostate for both the cases. B$_{\rm{H}}$, on the other hand, behave differently as internal spin configuration changes significantly during downsweep. As the bias field decreases from 150\,mT to 0\,mT, spins starts rotating to align itself along it's easy-axis viz., perpendicular to bias field orientation. As Field is further increased from 0\,mT to 150\,mT in (-)ve x direction, spin reorients itself along field direction by rotation. This results, in symmetric SW dispersion across 0\,mT field. Furthermore, as the structure with mixed coordinated CE-vertex carry a charge cloud around center vertex due to uncompensated charges at remanence, we observe a bifurcation at 0\,mT. However, for the even coordinated OE-vertex local charges appear at two corner vertices along the bias field orientation. This results in clear merging of SW modes and a shoulder peak.\\
Thus Detailed investigation performed for both geometries (see Fig.\ref{fig: Intro figure}) suggest that there is no perceptible difference on field dependence of spin wave dynamical behavior due to origin of local magnetic charge. However, SW modes excited in this sort of structures show significant mode tuning even by varying a single parameter viz., Bias field strength. This can further be tune by giving angle between easy axes and external magnetic field as an additional degree of freedom. Hence, the study suggest that these structures manifest itself as a potential candidate for magnonic crystals. Our simulation work on OE vertex with localized excitation suggests that appreciable directional propagation in preferential direction can be achieved in disconnected dipolar coupled nanomagnets by carefully tuning the location of local charges or excited type-III state and bias field orientation. This suggest that information embedded in form of SW phase and amplitude can be processed without the integration of current path which can be employ to resolve the persisting joule heating issue in cMOS based devices. Hence, all magnonic directional propagation behavior in studied sASI shows prospect in information processing and de-multiplexing devices. Further, response of SW in these structures can be utilized for fundamental studies such as the identification of higher order excitation e.g., monopole and anti-monopole from the CV mode. 
\begin{acknowledgements}
We acknowledge High Performance Computation (HPC) facility of IIT Delhi. N.A. is thankful to the Council of Scientific Industrial Research (CSIR), Government of India for research fellowship.
\end{acknowledgements}


\section*{References}
\begin{thebibliography}{99}
\bibitem{kruglyak2010preface}V. Kruglyak, S. Demokritov, and D. Grundler, Magnonics. J. Phys. D \textbf{43}, 260301 (2010).
\bibitem{neusser2009magnonics}S. Neusser and D. Grundler, Magnonics: spin waves on the nanoscale, Adv. Mater. \textbf{21}, 2927 (2009).
\bibitem{lenk2011building}B. Lenk, H. Ulrichs, F. Garbs, and M. M¨unzenberg, The building blocks of magnonics, Phys. Rep. \textbf{507}, 107 (2011).
\bibitem{demokritov2012magnonics}S. O. Demokritov and A. N. Slavin, Magnonics: From fundamentals to applications, Vol. 125 (Springer Science \& Business Media, 2012).
\bibitem{nikitov2015magnonics}S. A. Nikitov, D. V. Kalyabin, I. V. Lisenkov, A. N. Slavin, Y. N. Barabanenkov, S. A. Osokin, A. V. Sadovnikov, E. N. Beginin, M. A. Morozova, Y. P. Sharaevsky, et al., Magnonics: a new research area in spintronics and spin wave electronics, Physics-Uspekhi \textbf{58}, 1002 (2015).
\bibitem{liu2018organic}H. Liu, C. Zhang, H. Malissa, M. Groesbeck, M. Kavand, R. McLaughlin, S. Jamali, J. Hao, D. Sun, R. A. Davidson, \textit{et al.},  Organic-based  magnon  spintronics,  Nat. mat. \textbf{17}, 308 (2018).
\bibitem{au2012nanoscale}Y.  Au,  M.  Dvornik,  O.  Dmytriiev,  and  V.  Kruglyak, Nanoscale  spin  wave  valve  and  phase  shifter,  Appl. Phys. Lett. \textbf{100}, 172408 (2012).
\bibitem{abeed2020experimental}  M.  A.  Abeed  and  S.  Bandyopadhyay,  Experimental demonstration  of  an  extreme  subwavelength  nanomagnetic  acoustic  antenna  actuated  by  spin–orbit  torque from a heavy metal nanostrip, Adv. Mater. Technol. \textbf{5}, 1901076 (2020).
\bibitem{wang2018reconfigurable} Q. Wang, P. Pirro, R. Verba, A. Slavin, B. Hillebrands, and A. V. Chumak, Reconfigurable nanoscale spin-wave directional coupler, Sci. Adv. \textbf{4}, e1701517 (2018).
\bibitem{vogt2014realization} K. Vogt, F. Y. Fradin, J. E. Pearson, T. Sebastian, S. D.Bader, B. Hillebrands, A. Hoffmann, and H. Schultheiss, Realization of a spin-wave multiplexer, Nat. Commun. \textbf{5}, 1 (2014).
\bibitem{kostylev2005spin} M. Kostylev, A. Serga, T. Schneider, B. Leven, and, B. Hillebrands, Spin-wave logical gates, Appl. Phys. Lett. \textbf{87}, 153501 (2005).
\bibitem{papp2017nanoscale}A. Papp, W. Porod, ´ A. I. Csurgay, and G. Csaba, ´ Nanoscale spectrum analyzer based on spin-wave interference, Sci. Rep. \textbf{7}, 1 (2017).
\bibitem{chen2021reconfigurable}J. Chen, H. Wang, T. Hula, C. Liu, S. Liu, T. Liu, H. Jia, Q. Song, C. Guo, Y. Zhang, et al., Reconfigurable spinwave interferometer at the nanoscale,Nano Lett. \textbf{21}, 6237 (2021).
\bibitem{yu2013omnidirectional} H. Yu, G. Duerr, R. Huber, M. Bahr, T. Schwarze, F. Brandl, and D. Grundler, Omnidirectional spin-wave nanograting coupler, Nat. Commun. \textbf{4}, 1(2013).
\bibitem{grollier2016spintronic}J. Grollier, D. Querlioz, and M. D. Stiles, Spintronic nanodevices for bioinspired computing, Proc. IEEE \textbf{104}, 2024 (2016).
\bibitem{arava2018computational}H. Arava, P. M. Derlet, J. Vijayakumar, J. Cui, N. S. Bingham, A. Kleibert, and L. J. Heyderman, Computational logic with square rings of nanomagnets, Nanotechnology \textbf{29}, 265205 (2018).
\bibitem{arava2019engineering}H. Arava, N. Leo, D. Schildknecht, J. Cui, J. Vijayakumar, P. M. Derlet, A. Kleibert, and L. J. Heyderman, Engineering relaxation pathways in building blocks of artificial spin ice for computation, Phys. Rev. Appl. \textbf{11}, 054086 (2019).
\bibitem{caravelli2022artificial}F. Caravelli, G.-W. Chern, and C. Nisoli, Artificial spin ice phase-change memory resistors, New J. Phys. (2022).
\bibitem{torrejon2017neuromorphic} J. Torrejon, M. Riou, F. A. Araujo, S. Tsunegi, G. Khalsa, D. Querlioz, P. Bortolotti, V. Cros, K. Yakushiji, A. Fukushima, \textit{et al.,} Neuromorphic computing with nanoscale spintronic oscillators, Nature \textbf{547}, 428 (2017).
\bibitem{li2019strong} Y. Li, T. Polakovic, Y.L. Wang, J. Xu, S. Lendinez, Z. Zhang, J. Ding, T. Khaire, H. Saglam, R. Divan,
\textit{et al.,} Strong coupling between magnons and microwave
photons in on-chip ferromagnet-superconductor thin-film
devices, Phys. Rev. Lett. \textbf{123}, 107701 (2019).
\bibitem{kaffash2021nanomagnonics} M. T. Kaffash, S. Lendinez, and M. B. Jungfleisch, Nanomagnonics with artificial spin ice, Phys. Lett. A \textbf{402}, 127364 (2021).
\bibitem{lendinez2021observation} S. Lendinez, M. Taghipour Kaffash, and M. B. Jungfleisch, Observation of mode splitting in artificial spin ice: A comparative ferromagnetic resonance and brillouin light scattering study, Appl. Phys. Lett. \textbf{118}, 162407 (2021).
\bibitem{gartside2021reconfigurable} J. C. Gartside, A. Vanstone, T. Dion, K. D. Stenning, D. M. Arroo, H. Kurebayashi, and W. R. Branford, Reconfigurable magnonic mode-hybridisation and spectral control in a bicomponent artificial spin ice, Nat. Commun. \textbf{12}, 1 (2021).
\bibitem{caravelli2020logical} F. Caravelli and C. Nisoli, Logical gates embedding in artificial spin ice, New J. Phys. \textbf{22}, 103052
(2020).
\bibitem{iacocca2020tailoring} E. Iacocca, S. Gliga, and O. G. Heinonen, Tailoring spin wave channels in a reconfigurable artificial spin ice, Phys. Rev. Appl. \textbf{13}, 044047 (2020).
\bibitem{gliga2020dynamics} S. Gliga, E. Iacocca, and O. G. Heinonen, Dynamics of
reconfigurable artificial spin ice: Toward magnonic functional materials, APL Mater. \textbf{8}, 040911 (2020).
\bibitem{wang2006artificial}  R. Wang, C. Nisoli, R. S. d. Freitas, J. Li, W. McConville, B. Cooley, M. Lund, N. Samarth, C. Leighton, V. H. Crespi, \textit{et al.,} Artificial ‘spin ice’ in a geometrically frustrated lattice of nanoscale ferromagnetic islands, Nature \textbf{439}, 303 (2006).
\bibitem{morgan2011thermal} J. P. Morgan, A. Stein, S. Langridge, and C. H. Marrows, Thermal ground-state ordering and elementary excitations in artificial magnetic square ice, Nat. Phys. \textbf{7}, 75 (2011).
\bibitem{keswani2021controlled}N. Keswani, R. J. Lopes, Y. Nakajima, R. Singh, N. Chauhan, T. Som, D. S. Kumar, A. R. Pereira, and P. Das, Controlled creation and annihilation of isolated robust emergent magnetic monopole like charged vertices in square artificial spin ice, Sci. Rep. \textbf{11}, 1 (2021).
\bibitem{jungfleisch2017high} M. B. Jungfleisch, J. Sklenar, J. Ding, J. Park, J. E. Pearson, V. Novosad, P. Schiffer, and A. Hoffmann, High frequency dynamics modulated by collective magnetization reversal in artificial spin ice, Phys. Rev. Appl. \textbf{8}, 064026 (2017).
\bibitem{kapaklis2012melting} V. Kapaklis, U. B. Arnalds, A. Harman-Clarke, E. T. Papaioannou, M. Karimipour, P. Korelis, A. Taroni, P. C. Holdsworth, S. T. Bramwell, and B. Hj¨orvarsson, Melting artificial spin ice, New J. Phys. \textbf{14}, 035009 (2012).
\bibitem{qi2008direct}Y. Qi, T. Brintlinger, and J. Cumings, Direct observation of the ice rule in an artificial kagome spin ice, Phys. Rev. B \textbf{77}, 094418 (2008)
\bibitem{talapatra2021coupled} A. Talapatra and A. Adeyeye, Coupled magnetic nanostructures: Engineering lattice configurations, Appl. Phys. Lett. \textbf{118}, 172404 (2021).
\bibitem{chaurasiya2021comparison} A. K. Chaurasiya, A. K. Mondal, J. C. Gartside, K. D. Stenning, A. Vanstone, S. Barman, W. R. Branford, and A. Barman, Comparison of spin-wave modes in connected and disconnected artificial spin ice nanostructures using brillouin light scattering spectroscopy, ACS nano (2021).
\bibitem{saha2021spin}S. Saha, J. Zhou, K. Hofhuis, A. K´akay, V. Scagnoli, L. J. Heyderman, and S. Gliga, Spin-wave dynamics and symmetry breaking in an artificial spin ice, Nano Lett. \textbf{21}, 2382 (2021).
\bibitem{arora2021spin}N. Arora and P. Das, Spin wave spectral probing of degenerate microstates in building-block of square artificial spin ice, AIP Adv. \textbf{11}, 035030 (2021).
\bibitem{arroo2019sculpting}D. M. Arroo, J. C. Gartside, and W. R. Branford, Sculpting the spin-wave response of artificial spin ice via microstate selection, Phys. Rev. B \textbf{100}, 214425 (2019).
\bibitem{iacocca2016reconfigurable}E. Iacocca, S. Gliga, R. L. Stamps, and O. Heinonen, Reconfigurable wave band structure of an artificial square ice, Phys. Rev. B \textbf{93}, 134420 (2016).
\bibitem{montoncello2018mutual}F. Montoncello, L. Giovannini, W. Bang, J. Ketterson, M. Jungfleisch, A. Hoffmann, B. W. Farmer, and L. E. De Long, Mutual influence between macrospin reversal order and spin-wave dynamics in isolated artificial spinice vertices, Phys. Rev. B \textbf{97}, 014421 (2018).
\bibitem{vansteenkiste2014design} A. Vansteenkiste, J. Leliaert, M. Dvornik, M. Helsen, F. Garcia-Sanchez, and B. Van Waeyenberge, The design and verification of mumax3, AIP Adv. \textbf{4}, 107133 (2014).
\bibitem{buschow2003handbook} K. H. J. Buschow, \textit{Handbook of magnetic materials} (Elsevier, 2003).
\bibitem{barman2009dynamic} A. Barman and S. Barman, Dynamic dephasing of magnetization precession in arrays of thin magnetic elements, Phys. Rev. B \textbf{79}, 144415 (2009).
\bibitem{castelnovo2008magnetic} C. Castelnovo, R. Moessner, and S. L. Sondhi, Magnetic monopoles in spin ice, Nature \textbf{451}, 42 (2008).
\bibitem{keswani2020accessing}  N. Keswani, R. Singh, Y. Nakajima, T. Som, and P. Das, Accessing low-energy magnetic microstates in square artificial spin ice vertices of broken symmetry in static magnetic field, Phys. Rev. B . \textbf{102}, 224436 (2020).
\bibitem{chern2016magnetic} G.-W. Chern and P. Mellado, Magnetic monopole polarons in artificial spin ices, EPL \textbf{114}, 37004 (2016).
\bibitem{gilbert2014emergent}I. Gilbert, G.-W. Chern, S. Zhang, L. O’Brien, B. Fore, C. Nisoli, and P. Schiffer, Emergent ice rule and magnetic charge screening from vertex frustration in artificial spin ice, Nat. Phys. \textbf{10}, 670 (2014)
\end{thebibliography}
\end{document}